\documentclass[11pt]{article}
\def\be{\begin{equation}}
\def\ee{\end{equation}}
\def\ba{\begin{eqnarray}}
\def\ea{\end{eqnarray}}
\def\la{\langle}
\def\ra{\rangle}
\def\a{\alpha}
\def\b{\beta}

\def\h{\hskip 1cm}
\def\hh{\hskip 2cm}
\def\lo{\longrightarrow}
\usepackage{epsfig}
\begin{document}
\begin{titlepage}
\vspace{4cm}
\begin{center}{\Large \bf Entanglement and optimal strings of qubits for memory channels}\\
\vspace{2cm}V. Karimipour \footnote{Corresponding author,
email:vahid@sharif.edu}\h L.
Memarzadeh\footnote{email:laleh@mehr.sharif.edu}, \\
\vspace{1cm} Department of Physics,\\ Sharif University of Technology,\\
P.O. Box 11365-9161,\\ Tehran, Iran
\end{center}
\vskip 3cm
\begin{abstract}

We investigate the problem of enhancement of mutual information by
encoding classical data into entangled input states of arbitrary
length and show that while there is a threshold memory or
correlation parameter beyond which entangled states outperform the
separable states, resulting in a higher mutual information, this
memory threshold increases toward unity as the length of the string
increases. These observations imply that encoding classical data
into entangled states may not enhance the classical capacity of
quantum channels.

\end{abstract}
\end{titlepage}

\section{Introduction}
Consider a quantum channel defined by a completely positive trace
preserving map $\rho\lo \mathcal{E}^{(1)}(\rho)$. We use ${\mathcal
E}^{(1)}$ for describing the action of the channel on one qubit and
${\mathcal E}^{(n)}$ for its action on a string of qubits of length
$n$. By  encoding classical data into quantum states and performing
optimal measurements at the output, such a channel can be used for
communicating classical information. One is then faced with the
natural question of which states are the optimal ones for encoding
the input data, that is, which ensemble of input states maximize the
mutual information
between the sender and the receiver. \\

If the input strings are disentangled and if  consecutive uses of
the channel are not correlated to each other, that is for memoryless
channels, then the action of the channel on a string $\rho_1\otimes
\rho_2\cdots \otimes \rho_n$ is simple, namely it is given by
${\mathcal E}^{n}(\rho_1\otimes \rho_2\cdots
\otimes\rho_n)=\bigotimes_{i=1}^n {\mathcal E}^{1}(\rho_i)$. However
in general one may want to encode classical data into entangled
strings or, consecutive uses of the channel may be correlated to
each other, in which case ${\mathcal E}^{n}\ne {{\mathcal
E}^{1}}^{\otimes n}$. In these cases the output strings are
no longer simple functions of the input strings.\\

In these cases we are dealing with a strongly correlated quantum
system the correlations of which either result from the
entanglements of the input states, or from the memory of the channel
itself. And for this reason we should anticipate grave difficulties
in analytical tackling of the problem. Nevertheless we try to gain an insight by studying examples. \\

Given an ensemble of input states $\varepsilon:=\{\rho_i, p_i\}$
where $p_i$ are the probabilities of states $\rho_i$, and $\rho_i$
are states of $n-$ input qubits, the mutual information is defined
as

\begin{equation}\label{mutual}
    I_n(\varepsilon):=
    S({\mathcal E}^{(n)}(\sum_ip_i\rho_i))-\sum_{i}p_iS({\mathcal E}^{n}(\rho_i)),
\end{equation}

where $S(\rho)\equiv -tr(\rho \log \rho)$ is the von Neumann entropy
of a state $\rho$.\\

A basic question of information theory is whether there is any
advantage in using entangled states as input states, that is,
whether or not encoding the classical data into entangled rather
than separable states increases the mutual information. For the
case when multiple uses of the channel are not correlated, there
are partial evidence based on studying concrete examples
\cite{K1}, \cite{K2}, \cite{M1}that the optimal states are
separable and hence there is no advantage in
using entangled states.\\

However if multiple uses of the channel are correlated, then there
are pieces of evidence that entangled states become advantageous,
once the correlation exceeds a critical value. \\

In \cite{M2, M3} a Pauli channel with partial memory, was studied.
The action of the channel on two consecutive qubits is given by
the following map:
\begin{equation}\label{2q}
{\mathcal E}(\rho)=\sum_{i,j} P_{ij} (\sigma_i \otimes \sigma_j)\rho
(\sigma_i \otimes \sigma_j)^{\dag},
\end{equation}
where $P_{ij}$ denotes the probability of two consecutive errors
$\sigma_i,\sigma_j$  and is defined as:
\begin{equation}\label{Pij}
P_{ij}=(1-\mu)p_i p_j+\mu \delta_{i,j}p_i=[(1-\mu)p_i+\mu
\delta_{i,j}]p_j.
\end{equation}
Here $p_i$ is the probability of the error operator $\sigma_i$,
$i=0, 1,2,3$ on one single qubit. Thus with probability $\mu$ the
channel acts on the second qubit with the same error operator as on
the first qubit, and with probability $(1-\mu)$ it acts on the
second qubit independently, hence the name partial memory.\\

Physically the parameter $\mu$ is determined by the relaxation time
of the channel when a qubit passes through it. In order to remove
correlations, one can wait until the channel has relaxed to its
original state before sending the next qubit, however this lowers
the rate of information transfer. Thus it is necessary to consider
the performance of the channel for arbitrary values of $\mu$ to
reach a compromise between various factors which determine the final
rate of information
transfer.\\

In \cite{M2} it was shown by analytical arguments and numerical
searches in the space of two-qubit input states, that for the
depolarizing channel \cite{Nil} there is a sharp transition in the
type of optimal states from separable to maximally entangled states
when the memory parameter $\mu$ passes beyond a critical value
$\mu_c$. A similar result was established analytically by the same
authors in \cite{M3} who considered a particularly symmetric channel
with ($p_0=p_3$ and $p_1=p_2$). Inspired by these works, similar
results have been shown for generalized Pauli channels acting on
states of arbitrary dimensions or qudits in \cite{KM,Karpov} and for
bosonic Gaussian channels in \cite{C}.\\

However all the above studies have been restricted to strings of
states of length $n=2$.  As is well known from theorems on data
compression in classical \cite{Shannon} and quantum
\cite{Schumacher} information theory, one should encode classical
data into arbitrarily long sequences of bits or qubits. This is
quite necessary if one wants to encode with arbitrarily high
probability only typical sequences and achieve maximum compression
of data and minimum decoding errors.\\
This requirement is also reflected in the definition of classical
capacity of quantum channels which is given by
\begin{equation}\label{C}
C:=\lim_{n\rightarrow\infty} C_n
\end{equation}
where
\begin{equation}\label{CN}
    C_n:=\frac{1}{n}Sup_{\varepsilon}I_n(\varepsilon),
\end{equation}
in which $n$ is the length of input string of states.  \\

Thus the question of whether entanglement enhances the mutual
information or not should be addressed for strings of arbitrary
length and not just strings of length $2$. More concretely one may
ask if it is advantageous to encode $2^n$ bits of information into
completely separable states of $n$ qubits or else, into maximally
entangled states. Only then one can make precise statements as to
the enhancement effect of entanglement
on the mutual information and capacity of quantum channels. \\

We should stress that the answer to this question, whatever it may
be,  does not invalidate the previous results, that entangled states
give a higher mutual information than separable states, as long as
we use a correlated channel "twice". However if consecutive uses of
a channel are correlated, and we have to encode our data into
arbitrary long strings, that is we are streaming the data into the
channel, then we should consider the effect of this correlation on
all the qubits of the strings. Thus we are asking the question of
 "which states maximize the mutual information in the space of all states of $n-$
 qubits?"\\

Unfortunately answering this question in its full generality is
almost intractable, for at least two reasons. First it is an
extremely difficult task to optimize the mutual information over all
ensembles of $n$ qubit states, due to the exponentially large number
of parameters involved. Second we do not have good measures to
characterize various types of entanglement in multi-partite states.
For example in contrast to the two-party case in which we have only
one class of entangled states, for $n$-parties there are numerous
inequivalent classes of entangled states the number of
which grows very rapidly with the number of $n$ \cite{Cirac}. \\

Nevertheless we can gain an insight into this problem by comparing
only two types of ensembles, namely an ensemble of pure product
states and an ensemble of pure maximally entangled states like the
Greenberger-Horne-Zeilinger (GHZ) states. We should stress that even
in this case we are still faced with a strongly correlated quantum
many body system which poses many computational difficulties for its
solution.
\\
There are several reasons in favor of this restricted choice. First
following the work of \cite{M3} we will show in the sequel that the
problem of optimizing the mutual information over input ensembles
reduces for Pauli channels to finding a single pure state that
minimizes the output entropy. Second, previous examples \cite{M2,
M3, Karpov,KM} mentioned above show that as we vary the correlation
parameter of the channel $\mu$, the optimal input state (which
minimizes the output entropy) changes sharply from a separable state
to a maximally entangled state and for no value of this parameter a
state with an intermediate value of entanglement is optimal (We will
elaborate on this point later in the introduction). Finally an
analytic calculation of the output entropy which requires
diagnolization of a $2^n\times 2^n$ matrix is impossible for an
arbitrary $n-$ qubit pure state
containing many parameters.\\

Now for the elaboration mentioned above: there is not a single class
of maximally entangled states for arbitrary $n$. For example for
$n=3$ there are two inequivalent classes which can not be
transformed to each other by invertible local operations.  Lack of
knowledge of all these classes for arbitrary $n$ and the particular
simplicity of the $GHZ$ states compels us to consider only this
class analytically. In order to substantiate our arguments we also
consider other types of encoding of input states, like strings of
Bell states by numerical means. However due to the exponential
growth of the required time, we have been able only to consider
strings of a few number of Bell states which again lead to the same result as stated above.  \\

We have presented in figures (\ref{I}) and (\ref{J}) the entropy of
output states for several other types of encoding for strings
of length 3 and 4.\\
These figures clearly show that for $n=3$ the optimum input state
is a separable state and for $n=4$\\
the minimum output entropy state is separable for small value of
correlation and one or the other type of entangled states for high
value of correlation. However in all types of encodings the
threshold parameter increases with the length of the strings.\\

Does these results prove conclusively that encoding classical data
into entangled states can not enhance the capacity of quantum
channels? Certainly no, because we have not made an exhaustive
search over the space of all $n-$ party states.  Nevertheless our
result casts doubt on the previous hope that this type of
entanglement, namely encoding classical data into entangled input
states,  may increase the capacity of quantum channels for
transmitting classical data.\\

The structure of this paper is as follows: In section (\ref{out})
we consider the Pauli channel with partial memory and calculate
its output states when we pass through it a string of $n$-qubit in
either separable or GHZ form. In section (\ref{Entropy}) we
diagonalize the output states and find the eigenvalues of the
output states in these two cases to calculate the output entropies
and find the critical value of memory above which $GHZ$ states
take over the separable states in maximizing the mutual
information. \\
In section (\ref{other}) we consider other types of entangled states for encoding where we restrict ourselves to strings of length $n=3$ and $n=4$.\\
Finally section (\ref{D}) concludes the paper with a discussion and
summary of the results. Appendix A contains some details of
calculations.\\

\section{The passage of a string of qubits through a Pauli channel with
partial memory}\label{out}

\subsection{Action of the channel}
\begin{figure}
  \centering
  \epsfig{file=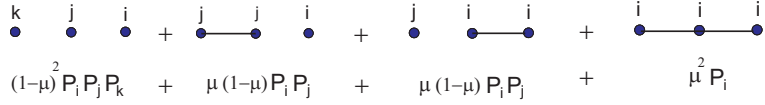,width=10cm}
  \caption{Various types of errors on a string of three qubits and their corresponding graphs. The rightmost graph represents an error which repeats for
  the second and third bits with probability $\mu^2$ and the leftmost graph represents three independent errors.}
  \label{error}
\end{figure}

The action of a Pauli channel with partial memory on a string of $n$
qubits is a natural generalization of equation (\ref{2q}). Before
considering the general case, it would be helpful to study the
simpler case of $n=3$. In this case we have
\begin{equation}\label{EN3}
{\mathcal E}^{(3)}(\rho)=\sum_{ijk} P_{ijk}
(\sigma_i\otimes\sigma_j\otimes\sigma_k)
\rho(\sigma_i\otimes\sigma_j\otimes\sigma_k)^{\dag}.
\end{equation}
The memory parameter $\mu$ is contained in the probabilities
$P_{ijk}$ which determine the probability of the errors
$\sigma_i\otimes\sigma_j\otimes\sigma_k$. Recalling that $(1-\mu)$
is the probability of independent errors on two consecutive qubits ,
and $\mu$ is the probability of identical errors, $P_{ijk}$ can be
written as follows:
\begin{equation}\label{Pijk}
P_{ijk}=[(1-\mu)p_i+\mu\delta_{i,j}][(1-\mu)p_j+\mu\delta_{j,k}]p_k.
\end{equation}
A good way to visualize the pattern of errors is via diagrams
depicted in figure (\ref{error}). A dot with a Latin index say $i$
on it represents an error $\sigma_i$ which happens with
probability $p_i$, and a line represent correlation between two
consecutive errors which happens with probability $\mu$. Thus the
pattern of errors for a string of three qubits is the one shown
in figure (\ref{error}) and is concisely represented in equation
(\ref{Pijk}).

Note from figure (\ref{error}) that the sum of probabilities of
all types of errors on three qubits adds to unity as we expect:
\begin{equation}\label{errors}
    \sum_{i,j,k}\left[(1-\mu)^2p_ip_jp_k+2\mu(1-\mu)p_ip_j+\mu^2p_i\right]=1.
\end{equation}
The action of the channel on a string of $n-$ qubits is given by
\begin{equation}\label{EN}
{\mathcal E}^{(n)}(\rho)=\sum_{i_1\cdots i_n=1}^3 P_{i_1\cdots
i_n}(\sigma_{i_1}\otimes\cdots\otimes
\sigma_{i_n})\rho(\sigma_{i_1}\otimes\cdots\otimes \sigma_{i
_n})^{\dag},
\end{equation}
where
\begin{equation}\label{PN}
P_{i_1\cdots i_n}=p_{i_n}\prod_{m=1}^{n-1}
\left[(1-\mu)p_{i_m}+\mu \delta_{i_m,i_{m+1}}\right].
\end{equation}
Thus in passing through the channel any two consecutive qubits
undergo random independent errors with probability $1-\mu$ and
identical (correlated) errors with probability $\mu$. This should be
the case if the channel has a memory depending on its relaxation
time and if we stream the qubits through it. \\

\subsection{Properties of the channel}
Since any two Pauli operators either commute or anti-commute with
each other, the channel has the following very important property
\be {\mathcal E}^{(n)}(U\rho U^{\dag})=U {\mathcal
E}^{(n)}(\rho)U^{\dag},\ee
where $U$ is the $n-$ fold tensor product of any combination of Pauli operators.\\
Therefore for any state $\rho$ and any operator $U$ of the above
form we have
\begin{equation}\label{U}
    S({\mathcal E}^{(n)}(U\rho U^{\dag})) = S({\mathcal E}^{(n)}(\rho)),
\end{equation}
where $S$ is the von-Neumann entropy.\\

Following the arguments of \cite{M3}, we now see that the mutual
information is saturated by an equiprobable ensemble of the form
\be\label{ensemble}\varepsilon=\{\rho_{i_1,i_2,\cdots
i_n}:=(\sigma_{i_1}\otimes \cdots
\sigma_{i_n})\rho^*(\sigma_{i_1}\otimes \cdots
\sigma_{i_n})^{\dag}\},\ee where $\rho^*$ is the state which
minimizes the output entropy. The reason is that the state
$\sum_{i_1,i_2,\cdots i_N}\rho_{i_1,i_2,\cdots i_N}$ commutes with
all the irreducible representation of the Pauli group on $n$ qubits
and hence is proportional to the identity. Thus for an equiprobable
ensemble  of the above type, the first term of (\ref{mutual}) is
maximized while the second term is
minimized due to the minimum output entropy of $\rho^*$. \\
In this way an upper bound for the mutual information is obtained,
namely
\begin{equation}\label{calf}
I_n({\mathcal E}^{(n)})\leq n-S({\mathcal E}^{(n)}(\rho^*)).
\end{equation}
Thus the problem of finding the optimal ensemble for maximizing the
mutual information reduces to finding the single state $\rho^*$
which minimizes the output entropy. Moreover this state can be taken
to be a pure state. To see this we again repeat the argument of
\cite{M3} for completeness: any state $\rho^*$ has a decomposition
into pure states given by $\rho^*=\sum_ip_i|\Psi_i\ra\la \Psi_i|$.
Concavity of $S$ entails that
$$S(\rho^*)\geq \sum_{i}
p_iS(|\Psi_i\ra)>S(|\Psi^*\ra),$$ where $|\Psi^*\ra$ is the state
with minimal entropy in the decomposition. However by definition of
$\rho^*$ as the state with minimum entropy we should have
$\rho^*=|\Psi^*\ra\la \Psi^*|$. This completes the proof that we
should only search for a single state to find the optimal ensemble.
\subsection{The output states of the channel}
\subsubsection{Strings of length $n=3$ }
For simplicity let us first consider strings of length $n=3$. We
consider a specific type of symmetric Pauli channel, one for which
$p_0=p_3=p$ and $p_1=p_2=q$, with $p+q=\frac{1}{2}$. This type of
channel was first considered in \cite{M3} (for two-qubit strings)
for which analytical calculations were shown to be possible due
to the extra symmetry ${\mathcal E}^{(2)}(\rho)={\mathcal
E}^{(2)}(\sigma_3\otimes
\sigma_3\rho\sigma_3\otimes \sigma_3)$. \\
Using the definition of the channel in equation (\ref{EN3}), a
straightforward but lengthy calculation gives the output states
relating to separable $|000\ra $ and GHZ
$\frac{1}{\sqrt{2}}(|000\ra+|111\ra)$ input states (see the Appendix
for the general case of arbitrary $n$). We find

\begin{equation}\label{out3}
{\mathcal
E}^{(3)}(|000\ra)=\sum_{\a\b\gamma=0}^1\tilde{P}_{\a\b\gamma}|\a\b\gamma\ra\la
\a\b\gamma|,
\end{equation}

and
\begin{equation}
{\mathcal E}^{(3)}(|GHZ\ra)
=\sum_{\a\b\gamma=0}^1\tilde{P}_{\a\b\gamma}(|\a\b\gamma\ra\la
\a\b\gamma|+|\overline{\a},\overline{\b},\overline{\gamma}\ra\la
\overline{\a},\overline{\b},\overline{\gamma}|),
\end{equation}
where $\overline{\a}=\a+1$ mod $2$, and
\begin{equation}\label{tildeP3}
\tilde{P}_{\a\b\gamma}=[(1-\mu)\eta_{\a}+\mu\delta_{\a,\b}][(1-\mu)\eta_{\b}+\mu\delta_{\b,\gamma}]\eta_{\gamma},
\end{equation}
with $\eta_0=2p$ and $\eta_1=2q$. Note that
$\tilde{P}_{\a\b\gamma}$ is structurally similar to $P_{ijk}$
(although their indices have different ranges), which help us in
writing the output states for strings of arbitrary length.\\

\subsubsection{Strings of arbitrary length}
The output state of the channel depends on whether the length of the
string is even or odd.  In the appendix we show that for general
length $n$, we have
\begin{eqnarray}\label{outN}
{\mathcal E}^{(n)}(|0\ra^{\otimes
n})&=&\sum_{\a=0}^{2^n-1}\tilde{P}_{\a}|\a\ra\la\a|,\cr {\mathcal
E}^{(n)}(|GHZ_n\ra)&=&\frac{1}{2}\sum_{\a=0}^{2^n-1}
\left[\tilde{P}_{\a}(|\a\ra\la\a|
+|\overline{\a}\ra\la\overline{\a}|)
+\tilde{Q}_{\a}(|\a\ra\la\overline{\a}|
+|\overline{\a}\ra\la\a|\right],
\end{eqnarray}
where $|\a\ra=|\a_1,\a_2\cdots\a_n\ra$,

\begin{equation}
\tilde{P}_{\a}=\eta_{{\a}_n}\prod_{i=1}^{n-1}[(1-\mu)\eta_{{\a}_i}
+\mu\delta_{\a_i,\a_{i+1}}],
\end{equation}
and
\begin{equation}
\tilde{Q}_{\a}=\left\lbrace
\begin{array}{l}
\mu^{\frac{n}{2}}\eta_{{\a}_n}\delta_{\a_{n-1},\a_{n}}
\prod_{i=1}^{\frac{n}{2}-1}\delta_{\a_{2i-1},\a_{2i}}\left[(1-\mu)\eta_{{\a}_{2i}}
+\mu\delta_{\a_{2i},\a_{2i+1}}\right],\hskip 1cm  n=even\\
\\
0, \hskip 10.3cm n=odd.\\
\end{array}\right.
\end{equation}

\section{The entropies of the output states}\label{Entropy}
In this section we find the analytic expressions for the eigenvalues
of the output density matrices, when the input states are
respectively the separable
$|\mathbf{0}\ra$ and the GHZ state $|GHZ\ra$. \\
Since the output state of the separable state is diagonal, its
eigenvalues are simply given by $ \lambda_{\a}=\tilde{P}_\a$.\\

\begin{figure}
  \centering
  \epsfig{file=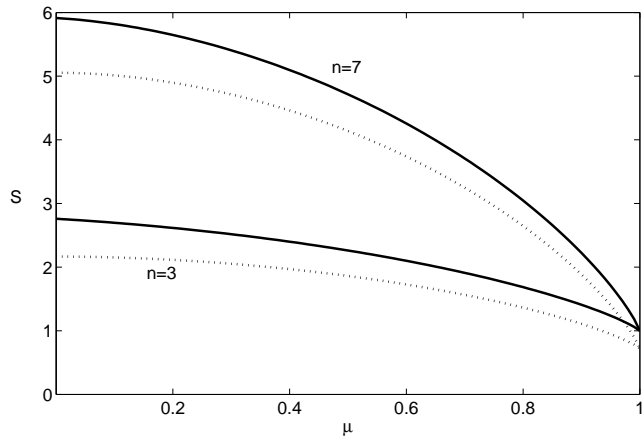,width=10cm}
  \caption{The output entropy of separable(...) and GHZ(---)
  input states with size $n=3$ and $n=7$, as a function of memory parameter $\mu$ in p=0.4.
   It seems that entangled states do not have any advantage over separable states for strings of odd length.}
  \label{odd}
\end{figure}

To find the eigenvalues of ${\mathcal E}(|GHZ\ra)$ we note that this
state is as embedding of blocks of the form:
\begin{equation}
\frac{1}{2}\left(\begin{array}{cc}
  \tilde{P}_{\a}+\tilde{P}_{\overline{\a}}&\tilde{Q}_{\a}+\tilde{Q}_{\overline{\a}} \\
  \tilde{Q}_{\a}+\tilde{Q}_{\overline{\a}} & \tilde{P}_{\a}+\tilde{P}_{\overline{\a}}
\end{array}\right),
\end{equation}
in different non-overlapping positions of a matrix.  Thus the
eigenvalues of the output state $\mathcal{E}(|GHZ\ra)$, are a
collection of the eigenvalues of these block. Thus we find the final
form of eigenvalues for both input states:
\begin{eqnarray}\label{lGHZ}
\lambda_{\a}&=& \tilde{P}_{\a}\hh \hh \hh for\ \ \ \ |\mathbf{0}\ra,
\cr
\lambda_{\a}&=&\frac{1}{2}(\tilde{P}_{\a}+\tilde{P}_{\overline{\a}}
\pm(\tilde{Q}_{\a}+\tilde{Q}_{\overline{\a}})) \hh \ \ for \ \ \ \
|GHZ\ra.
\end{eqnarray}

\begin{figure}
  \centering
  \epsfig{file=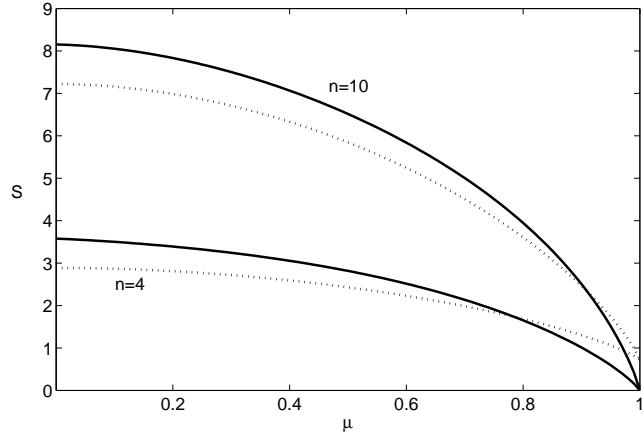,width=10cm}
  \caption{The output entropy of separable (...) and GHZ(---)
  input states with size $n=4$ and $n=10$, as a function of memory parameter $\mu$ for $p=0.4$.
  Entangled states seems to have an advantage for strings of even length.}
  \label{even}
\end{figure}

\begin{figure}
  \centering
  \epsfig{file=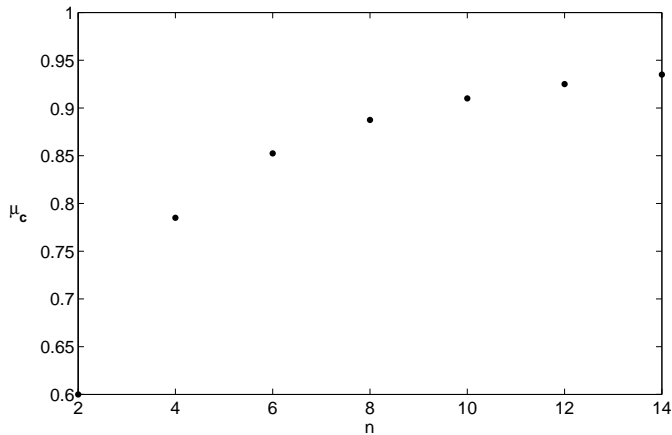,width=10cm}
  \caption{Critical memory as a function of the size of the input string for $p=0.4$.}
  \label{critical04}
\end{figure}

The subscript, $\a=\a_1\a_2\cdots\a_N$, range over the the numbers
$\{0,\ 2^n-1\}$ for separable states. For the GHZ states the same
range of subscripts produces the eigenvalues twice, since $\a$ and
$\overline{\a}$ lead to the same eigenvalues. So the calculated
entropy for the whole range of $\a\in \{0,2^n-1\}$ should be halved
to
obtain the correct  final value. \\
From these eigenvalues we can calculate the output entropies
$$S_{out}=-\sum_{\a}\lambda_{\a} \log \lambda_{\a}$$ as functions of
$p$ and $\mu$. For small values of $n$ the entropy can be evaluated
in closed form. However as we increase $n$ the number of eigenvalues
increases exponentially and a closed expression is not possible.
Figures (\ref{odd}) and (\ref{even}) show the output entropy for the
separable and GHZ states for strings of different length as a
function of the memory parameter $\mu$. It is seen that for odd
length of the string, the separable states are better than the GHZ
states for all values of the memory parameters. However for strings
of even-length the GHZ states are better once the memory parameter
passes a critical value. Figure (\ref{critical04}) shows the value
of this critical memory as a function of the length of the string
for a typical value of the error parameter. It is seen that as the
length of the string increases this critical value increases toward
unity. Thus for very large strings we conclude that entangled states
loose their advantage over
separable states altogether.\\

\section{Other types of encoding}\label{other}
\begin{figure}
  \centering
  \epsfig{file=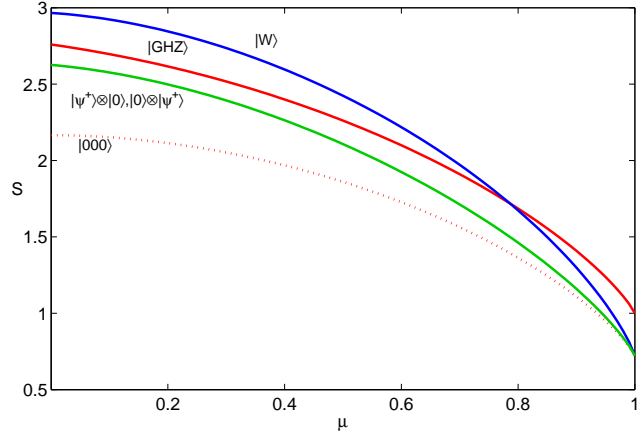,width=10cm}
  \caption{The output entropy for different entangled input strings of length n=3,\ and
  $p=0.4$.
   Separable states are better than other states for strings of odd length.}
  \label{I}
\end{figure}

\begin{figure}
  \centering
  \epsfig{file=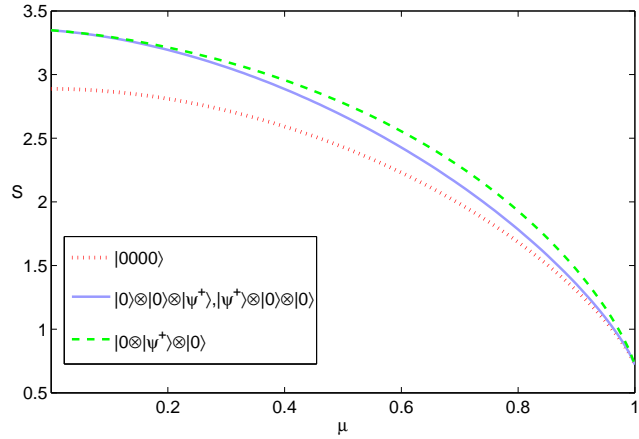,width=10cm}
  \caption{The output entropy for different entangled input strings of length n=4,\ and
  $p=0.4$.
  Again completely separable states have lower entropy than partially entangled
  strings.}
  \label{J}
\end{figure}

\begin{figure}
  \centering
  \epsfig{file=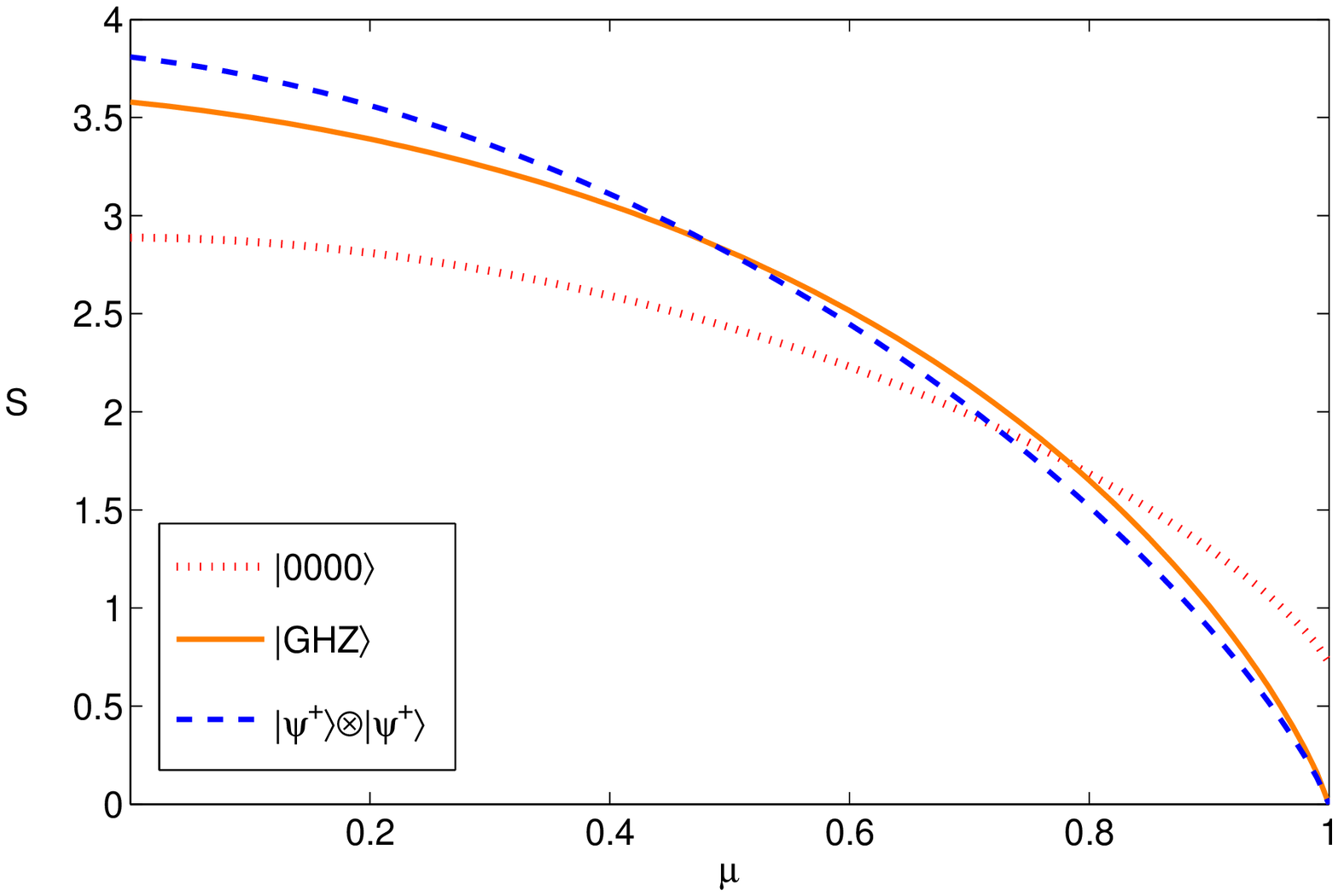,width=10cm}
  \caption{The output entropy for different entangled input strings of length n=4,\ and $p=0.4$. For high value of memory GHZ and pairs of Bell states
  take over the separable strings.}
  \label{K}
\end{figure}

In order to substantiate our arguments, we depict in figures
(\ref{I} ) and (\ref{J}) the output entropy for several types of
other states for strings of length $n=3$ and $n=4$. Each of these
states gives rise
to an equi-probable ensemble as defined in (\ref{ensemble}).\\
The calculations for these types of states have been done as
outlined in previous sections. For $n=3$ we have considered the
additional states $$|\psi^+\ra\otimes |0\ra,\h {\rm and}\h   |0\ra
\otimes |\psi^+\ra$$ where $
|\psi^+\ra=\frac{1}{\sqrt{2}}(|00\ra+|11\ra)$ is a Bell state and
a state of the form
$$|W\ra=\frac{1}{2}(|100\ra+|010\ra+|001\ra+|111\ra).$$ For $n=4$ we
have considered the additional states $$|\psi^+\ra\otimes
|\psi^+\ra,\ \ \ \ \ |\psi^+\ra\otimes |00\ra,\ \ \ \ |0\ra\otimes
|\psi^+\ra\otimes |0\ra \ \ \ {\rm and}\ \ \
 |00\ra\otimes |\psi^+\ra.$$\\
 Figure (\ref{I}) shows that for $n=3$ separable states are optimal for all values of $\mu$ while figures (\ref{J}) and (\ref{K}) show
 that the optimal state changes from a separable state to an entangled state as we increase $\mu$. It is
 seen
 that
 a only a
 string of two Bell states, or a GHZ state have a lower entropy then the separable state and other forms of partially entangled
 states are not better than separable states. Moreover it is seen that when the memory parameter increases a string of two Bell
 states is
 slightly better than the GHZ state.\\

\section{Summary}\label{D}

We have compared the mutual information for a memory Pauli channel
for two ensemble of input states of arbitrary length, a separable
ensemble and an ensemble constructed from the entangled GHZ
states. Our results show that for odd lengths of the strings,
separable states lead to a higher mutual information. However for
strings of even length GHZ states outperform the separable states
when the memory of the channel exceeds a critical value. The
value of this critical memory increases toward unity as the
length of the string increases implying that for arbitrarily long
sequences still the separable and GHZ states  perform equally
well. \\ Although previous works \cite{M2, M3,KM,Karpov,C} have
shown that entangled states are advantageous for encoding
classical data into two qubits, our results imply that this
advantage may not lead to a higher rate of information transfer,
since general theorems of information theory require that only
typical sequences be encoded and these should be encoded into
arbitrary large sequences for which we have shown that this
advantage no longer exists.

\section{Appendix A}
Here we briefly explain how to calculate the output states of the
channel. The derivation is simplified if we relabel the Pauli
error operators and the corresponding probabilities by two
indices instead of one. Thus the Pauli operators, including the
identity operator $I$ are denoted by $\sigma_{_{\a,\b}}$ where
$\a,\b=0,1$. Such an error operator acts with probability
$p_{_{\a,\b}}$.  We have $\sigma_{0,0}=I,\ \
\sigma_{0,1}=\sigma_z,\ \ \sigma_{1,0}=\sigma_x, $ and
$\sigma_{1,1}=i\sigma_y$ whose actions are compactly written as
\begin{equation}\label{sigma} \sigma_{_{\a,\b}}=\sum_{k=0}^1
(-1)^{\b k}|k+\a\ra\la k|. \hskip 1cm \a,\b=0,1.
\end{equation}
The channel acts on a single qubit as:
\begin{equation}
{\mathcal
E}^{(1)}(\rho)=\sum_{\a,\b=0}^1p_{_{\a,\b}}\sigma_{_{\a\b}}\rho
\sigma_{_{\a\b}}^{\dagger}.
\end{equation}

From (\ref{EN}) and (\ref{PN}), it's action on a string of $n$
qubits, is given by :
\begin{equation}
{\mathcal
E}^{(n)}(\rho)=\sum_{\mathbf{\a},\mathbf{\b}=0}^1P_{_{\mathbf{\a},\mathbf{\b}}}
\chi_{_{\mathbf{\a},\mathbf{\b}}}(\rho),
\end{equation}
in which $\mathbf{\a}=\a_1\a_2\cdots\a_n$,
$\mathbf{\b}=\b_1\b_2\cdots\b_n$, and
\begin{equation}\label{chi}
\chi_{_{\mathbf{\a},\mathbf{\b}}}(\rho)=
(\sigma_{_{\a_1\b_1}}\otimes\sigma_{_{\a_2\b_2}}\cdots\otimes\sigma_{_{\a_n\b_n}})\rho
(\sigma_{_{\a_1\b_1}}\otimes\sigma_{_{\a_2\b_2}}\cdots\otimes\sigma_{_{\a_n\b_n}})^{\dag}.
\end{equation}
and
\begin{equation}
P_{_{\mathbf{\a},\mathbf{\b}}}=\prod_{i=1}^{n-1}\left[(1-\mu)p_{_{\a_i\b_i}}+\mu\delta_{_{\a_i\a_{i+1}}}
\delta_{_{\b_i\b_{i+1}}}\right]p_{_{\a_n\b_n}}.
\end{equation}
It is now easy to calculate the output states of separable and GHZ
input states:
\begin{equation}\label{en}
{\mathcal
 E}^{(n)}(|\mathbf{0}\ra\la\mathbf{0}|)=\sum_{\mathbf{\a},\mathbf{\b}=0}^1P_{_{\mathbf{\a},\mathbf{\b}}}
\chi_{_{\mathbf{\a},\mathbf{\b}}}(|\mathbf{0}\ra\la\mathbf{0}|)
\end{equation}
and
\begin{equation}
{\mathcal
 E}^{(n)}(|GHZ\ra)=\frac{1}{2}\sum_{\mathbf{\a},\mathbf{\b}=0}^1P_{_{\mathbf{\a},\mathbf{\b}}}\left[
\chi_{_{\mathbf{\a},\mathbf{\b}}}(|\mathbf{0}\ra\la\mathbf{0}|)+\chi_{_{\mathbf{\a},\mathbf{\b}}}(|\mathbf{0}\ra\la\mathbf{1}|)+
\chi_{_{\mathbf{\a},\mathbf{\b}}}(|\mathbf{1}\ra\la\mathbf{0}|)+\chi_{_{\mathbf{\a},\mathbf{\b}}}(|\mathbf{1}\ra\la\mathbf{1}|)\right].
\end{equation}
From (\ref{chi}) and (\ref{sigma})  we have:
\begin{eqnarray}
\chi_{\mathbf{\a},\mathbf{\b}}(|\mathbf{0}\ra\la\mathbf{0}|)&=&|\mathbf{\a}\ra\la\mathbf{\a}|\cr
\chi_{\mathbf{\a},\mathbf{\b}}(|\mathbf{1}\ra\la\mathbf{1}|)&=&|\overline{\mathbf{\a}}\ra\la\overline{\mathbf{\a}}|\cr
\chi_{\mathbf{\a},\mathbf{\b}}(|\mathbf{0}\ra\la\mathbf{1}|)&=&|\mathbf{\a}\ra\la\overline{\mathbf{\a}}|(-1)^{\sum_{i=1}^n
\b_i }\cr
\chi_{\mathbf{\a},\mathbf{\b}}(|\mathbf{1}\ra\la\mathbf{0}|)&=&|\overline{\mathbf{\a}}\ra\la\mathbf{\a}|(-1)^{\sum_{i=1}^n
\b_i },
\end{eqnarray}
in which $\overline{\a}=\a+1$ mod 2. Combining these with
(\ref{en}) the output states are found to be:
\begin{eqnarray}
{\mathcal
 E}^{(n)}(|\mathbf{0}\ra\la\mathbf{0}|)&=&\sum_{\mathbf{\a},\mathbf{\b}=0}^1P_{\mathbf{\a},\mathbf{\b}}
|\mathbf{\a}\ra\la\mathbf{\a}|,\cr {\mathcal
 E}^{(n)}(|GHZ\ra)&=&\frac{1}{2}\sum_{\mathbf{\a},\mathbf{\b}=0}^1P_{\mathbf{\a},\mathbf{\b}}
(|\mathbf{\a}\ra\la\mathbf{\a}|+|\overline{\mathbf{\a}}\ra\la\overline{\mathbf{\a}}|)\cr
&+&\sum_{\mathbf{\a},\mathbf{\b}=0}^1P_{\mathbf{\a},\mathbf{\b}}
(-1)^{\sum_{n=1}^N\b_n}(|\mathbf{\a}\ra\la\overline{\mathbf{\a}}|+|\overline{\mathbf{\a}}\ra\la\mathbf{\a}|).
\end{eqnarray}
If we define $\widetilde{P}_{\mathbf{\a}}$, and
$\widetilde{Q}_{\mathbf{\a}}$ as follows:
\begin{equation}
\widetilde{P}_{\mathbf{\a}}=\sum_{\mathbf{\b}=0}^{2^n-1}P_{\mathbf{\a},\mathbf{\b}}\hskip
2cm \widetilde{Q}_{\mathbf{\a}}=\sum_{\mathbf{\b}=0}^{2^n-1}
P_{\mathbf{\a},\mathbf{\b}}(-1)^{\sum_{i=1}^n\b_i},
\end{equation}
the output states take the simple form:
\begin{eqnarray}
{\mathcal
 E}^{(n)}(|\mathbf{0}\ra\la\mathbf{0}|)&=&\sum_{\mathbf{\a}=0}^{2^n-1}\widetilde{P}_{\mathbf{\a}}
|\mathbf{\a}\ra\la\mathbf{\a}|,\cr {\mathcal
 E}^{(n)}(|GHZ\ra)&=&\frac{1}{2}\sum_{\mathbf{\a}=0}^{2^n-1}\left[
\widetilde{P}_{\mathbf{\a}}(|\mathbf{\a}\ra\la\mathbf{\a}|+|\overline{\mathbf{\a}}\ra\la\overline{\mathbf{\a}}|)
+\widetilde{Q}_{\mathbf{\a}}(|\mathbf{\a}\ra\la\overline{\mathbf{\a}}|+|\overline{\mathbf{\a}}\ra\la\mathbf{\a}|)\right].
\end{eqnarray}

For a symmetric Pauli channel in which $p_{00}=p_{01}=p$,
$p_{10}=p_{11}=q$ and $p+q=\frac{1}{2}$ the expressions of
$\tilde{P}_{\a}$ and $\tilde{Q}_{\a}$ are simplified to
\begin{equation}
\tilde{P}_{\a}=\eta_{{\a}_n}\prod_{i=1}^{n-1}\left[(1-\mu)\eta_{{\a}_i}
+\mu\delta_{\a_i,\a_{i+1}}\right],
\end{equation}
and
\begin{equation}
\tilde{Q}_{\a}=\left\lbrace
\begin{array}{l}
\mu^{\frac{n}{2}}\eta_{{\a}_n}\delta_{\a_{n-1},\a_{n}}
\prod_{i=1}^{\frac{n}{2}-1}\delta_{\a_{2i-1},\a_{2i}}\left[(1-\mu)\eta_{{\a}_{2i}}
+\mu\delta_{\a_{2i},\a_{2i+1}}\right],\ \ \ n=even\\
\\
0 \hh\hh \hh\hh\h\ \ , \ \ \ \ n=odd\\
\end{array}\right.
\end{equation}
where $\eta_0=2p$ and $\eta_1=2q$.\\

\begin{thebibliography}{}
\bibitem{K1}C. King and  M.B. Ruskai, IEEE Trans. Inf. Theory
            \textbf{47},192(2001).
\bibitem{K2}C. King, quant-ph/0103156
\bibitem{M1}D. Bru\ss, L. Faoro, C. Macchiavello, and G.M.
            Palma, J. Mod. Opt. \textbf{47}, 325, 2000.

\bibitem{M2}C. Macchiavello, and G.M. Palma, Phys. Rev. A \textbf{65},
            050301(R)(2002).
\bibitem{M3}C. Macchiavello, G.M. Palma, S. Virmani, Phys. Rev. A {\bf 69}, 010303,
(2004).
\bibitem{Nil}M. A. Nielsen, and I. L. Chuang;{\it{Quantum computation and quantum
information}},Cambridge University Press, Cambridge, 2000.
\bibitem{KM},V. Karimipour, L. Memarzadeh, Transition behavior in the capacity of correlated-noisy
channels in arbitrary dimensions, quant-ph/0603223, Phys. Rev. A, in
press.
\bibitem{Karpov}E. Karpov, D. Daems, N. J. Cerf, quant-ph/0603286
\bibitem{C}N.J. Cerf, J. Clavareau, C. Macchiavello. and J. Roland, Phys. Rev. A {\bf 72}, 042330
(2005).
\bibitem{Shannon}C. E. Shannon; {\it{A mathematical theory of
communication}}, Bell System Tech.J. \textbf{27}, 379-423,
623-656(1948); C. E. Shannon, W. Weavwe, \textit{The mathematical
theory of communication}, University of Illinois, Urbana(1949)
\bibitem{Schumacher}B. Schumacher and M. D. Westmoreland, Phys.Rev. A \textbf{56}, 131-138 (1997).
\bibitem{Cirac}W. D\"{u}r, G. Vidal, J. I. Cirac, Phys. Rev. A \textbf{62}, 062314 (2000)

\end{thebibliography}
\end{document}